\documentclass{article} 
\usepackage{amsmath}
\usepackage{amsfonts,amssymb}
\newtheorem{theorem}{Theorem}
\newtheorem{lemma}{Lemma}

\renewcommand{\H}{\mathcal{H}} 
\newcommand{\A}{\mathcal{A}} 
\newcommand{\B}{\mathcal{B}}
\newcommand{\Rl}{\mathbb{R}}

\newcommand{\spec}{\mathop{\rm spec}}
\newcommand{\be}{\begin{equation}}
\newcommand{\ee}{\end{equation}}
\newcommand{\bea}{\begin{eqnarray}}
\newcommand{\eea}{\end{eqnarray}}
\newcommand{\beann}{\begin{eqnarray*}}
\newcommand{\eeann}{\end{eqnarray*}}
\newcommand{\eq}[1]{(\ref{#1})}
\newcommand{\qed}{$\square$}
\begin{document}
\title{Lieb-Robinson Bounds and the\\[5pt] Exponential Clustering Theorem}
\author{Bruno Nachtergaele and Robert Sims\\[10pt]
Department of Mathematics\\ University of California at Davis\\
Davis CA 95616, USA\\
Email: bxn@math.ucdavis.edu, rjsims@math.ucdavis.edu}
\date{Version: August 12, 2005}
\maketitle  
\begin{abstract}
We give a Lieb-Robinson bound for the group velocity of a large class
of discrete quantum systems which can be used to prove that a non-vanishing
spectral gap implies exponential clustering in the ground state of
such systems.   
\end{abstract}
\renewcommand{\thefootnote}{$ $}
\footnotetext{Copyright \copyright\ 2005 by the authors. This article may be
reproduced in its entirety for non-commercial purposes.}
\section{Introduction}\label{sec:intro}

One of the folk theorems in quantum lattice models claims the equivalence
of the existence of a nonvanishing spectral gap and exponential decay of 
spatial correlations in the ground state. It has been known for some time
that there are exceptions to one direction of this equivalence. There 
are models with a unique ground state with exponential decay of
correlations but without a gap in the excitation spectrum above the ground
state. For a simple example see {Example 2} in \cite[p 596]{nachtergaele1996}.
In this paper we provide a rigorous proof of the other implication: a spectral
gap implies exponential decay in the ground state.   

In relativistic quantum field theory it has been known for a long
time that the existence of a mass (energy) gap implies exponential decay 
of spatial correlations. For example, in \cite{fredenhagen1985} Fredenhagen 
proves a general theorem applicable to arbitrary strictly local theories.
It is the strict locality, i.e., the fact that space-like separated 
observables commute, not the relativistic invariance per se, 
which plays a crucial role in the proof of exponential decay.
Non-relativistic models of statistical mechanics do not have strict locality,
but there is a finite
speed of propagation up to exponentially small corrections. This was first
proven by Lieb and Robinson \cite{lieb1972}. It is not a surprise
that the Lieb-Robinson bound can replace the strict locality property.
In particular, Wreszinski relied on it to prove a Goldstone Theorem in non-relativistic 
quantum statistical mechanics \cite{wreszinski1987}. Precisely how
to apply the Lieb-Robinson result to work around non-locality is not entirely 
obvious. Only recently, Hastings used it to derive
exponential clustering for lattice models with a gap \cite{hastings2004b}, 
and to obtain a generalization to higher dimensions of the
Lieb-Schultz-Mattis theorem \cite{lieb1961} in \cite{hastings2004a}. 
Our proof of exponential clustering is a rigorous version of Hastings' 
argument.

In the proof by Lieb and Robinson the lattice
structure played an essential role through the use of the Fourier  transform.
This was emphasized in a later version of their proof in  \cite{bratteli1997}.
Since lately there has been some interest in models  without translation
invariance or even without an underlying lattice structure such as 
spaces of fractal dimension \cite{dhar1977,tasaki1987,koma1995} or the
so-called complex networks \cite{albert2002,hastings2003}, 
we revisited the Lieb-Robinson result and provide here a proof that works 
for quite arbitrary models defined on a set of vertices with a metric. 

\section{Main Results}\label{sec:main}

We will consider quantum systems defined on a set of vertices $V$
with a finite dimensional Hilbert space $\H_x$ at each vertex $x\in V$.
At first we will assume that $V$ is finite.
For $X\subset V$, the Hilbert space associated with $X$ is the tensor
product $\H_X=\bigotimes_{x\in V}\H_x$, and the algebra of observables in $X$
is denoted by $\A_X=\B(\H_X)$. 
An interaction for such a system is a map $\Phi$ from the set of subsets
of $V$ to $\A_V$ such that $\Phi(X)\in\A_X$ and $\Phi(X) = \Phi(X)^*$
for all $X\subset V$. The Hamiltonian is defined by 
$$
H=\sum_{X\subset V} \Phi(X).
$$
The dynamics of the model is the one-parameter group of automorphisms,
$\{\tau_t\}_{t\in\Rl}$, defined by
$$
\tau_t(A)=e^{itH} A e^{-itH}, \quad A\in \A_V.
$$

We will assume that $V$ is equipped with a metric $d$. In the most common cases
$V$ is a graph, and the metric is given by the graph distance, $d(x,y)$,
which may be the length of the shortest path of edges connecting $x$ and $y$ in the 
graph. We will not, however, require an underlying graph structure for our
results. In terms of $d$ we define the diameter, $D(X)$, of a subset 
$X\subset V$ by
$$
D(X)=\max \{ d(x,y)\mid x,y\in X\}.
$$

Infinite systems can be introduced by considering a net of finite systems 
indexed
by finite sets $V$ partially ordered by inclusion. 
The $C^*$-algebra of observables, $\A$, is the norm completion of the 
union of the local observable algebras $\A_V$.
We will assume that there is a uniform bound, $N$, on the 
dimension of the single site
Hilbert spaces. An interaction $\Phi$ is defined as before but it is
necessary to impose a boundedness condition in order for the
finite-volume dynamics to converge to a strongly continuous
one-parameter group of automorphisms on $\A$. 
A standard reference for these infinite volume techniques
is \cite{bratteli1997}. The strength of a given interaction will be 
measured by a norm, $\Vert \cdot \Vert_\lambda$,
which for $\lambda >0$ is defined by
\be
\Vert \Phi \Vert_{\lambda} \, := \, \sup_{x \in V} \, \sum_{X \ni x}
\, |X|  \, \Vert \Phi(X) \Vert\, N^{2|X|} \, e^{ \lambda D(X)} \:
\ee
Here $\vert X\vert$ denotes the cardinality of the finite set $X$. For
finite $V$, the supremum in this definition is of course a
maximum. For infinite systems, finiteness of this norm is stronger
than what is required for the existence of the dynamics, but it is 
this norm that appears in the Lieb-Robinson 
bound \cite{lieb1972}. We will denote by $\B_\lambda$ the set of all 
potentials for the system under consideration such that 
$\Vert \Phi \Vert_\lambda < \infty$.
 
A Lieb-Robinson bound is an estimate for the following quantities:
\begin{equation} \label{CAB}
C_{A,B}(x,t) \, := \, \left[ \, \tau_t(A) \, , \, B \, \right],
\end{equation}
where $x\in V$, $t\in\Rl$, $A \in \A_x$, $B \in \A$. 
Due to the automorphism property of $\tau_t$, there is a symmetry
in the norm of such commutators:
$\Vert [\tau_t(A),B]\Vert = \Vert [A,\tau_{-t}(B)]\Vert $. 

It will be useful to consider
\begin{equation} \label{CB}
C_B(x,t) :=  \sup_{A \in\A_{x}} \frac{ \| C_{A,B}(x,t) \|}{ \| A \|}.
\end{equation}

In a typical application we would have $B\in\A_Y$, for some $Y\subset V$, 
and $x\in V\setminus Y$. Then, $A$ and $B$ commute and $C_B(x,0)=0$. 
A Lieb-Robinson bound then aims to show that
$C_B(x,t)$ is small for $|t|\leq T$ with $T$ proportional to the 
distance between $x$ and $Y$.

\begin{theorem}[Lieb-Robinson Bound]\label{thm:lr}
Fix $\lambda >0$, then for all $\Phi\in\B_\lambda$, $x\in
V$, $t\in \Rl$, and $B\in\A$, we have the bound
\be
C_B(x,t) \, \leq \, e^{2\, |t| \, \| \Phi \|_{\lambda}} C_B(x,0)
+  \sum_{y \in V: y \neq x} \,e^{- \, \lambda \,
  d(x,y)}  \left( e^{2 \, |t| \, \| \Phi \|_{\lambda}} - 1 \right)C_B(y,0)
\label{lrb}\ee
\end{theorem}

It is straightforward to derive from Theorem \ref{thm:lr} a bound for 
$\Vert [\tau_t(A),B]\Vert$ for general local observables $A\in\A_X$. One gets
$$
\Vert [\tau_t(A),B]\Vert\leq N^{2\vert X\vert}\Vert A\Vert\sum_{x\in X} 
C_B(x,t)\quad.
$$

For observables with overlapping supports, the Lieb-Robinson bound may
not be useful in the sense that the trivial bound $\Vert
[\tau_t(A),B]\Vert\leq 2 \Vert A\Vert\,\Vert B\Vert$
may be better. The problem of estimating $\Vert [\tau_t(A),B]\Vert$ for 
large $t$, in cases where it is expected to decay, is a separate 
issue that we do not address here.

In the case of observables with widely separated supports the information in \eq{lrb} 
is essentially equivalent to
\be
C_B(x,t) \, \leq \,  \sum_{y \in V} \, e^{2 \, |t| \, \| 
\Phi \|_{\lambda}- \, \lambda \,
  d(x,y)} C_B(y,0)
\label{lrb2}\ee

For strictly local $B\in\A_Y$, we can easily derive a more explicit bound
by using that, $C_B(y,0)\leq 2\Vert B \Vert \chi_Y(y)$, where
$\chi_Y$ is the characteristic function of the set $Y$. One obtains
for all $A\in \A_x$ and $B\in \A_Y$ the bound
\be
\Vert [\tau_t(A),B]\Vert\leq 2\vert Y\vert \, \Vert A\Vert \Vert B\Vert 
e^{2 \, |t| \, \| \Phi \|_{\lambda}  - \, \lambda \, d(x,Y)} 
\label{explicit_lr}\ee
If $x\not\in Y$, a stronger inequality holds:
\be
\Vert [\tau_t(A),B]\Vert\leq 2\vert Y\vert \, \Vert A\Vert \Vert B\Vert 
\left(e^{2 \, |t| \, \| \Phi \|_{\lambda}} -1\right) e^{- \, \lambda \, d(x,Y)} 
\label{beast}\ee

It would be interesting to have an analogue of \eq{beast} for complex
times. We may apply Theorem \ref{thm:lr} with $B$ replaced by $\tau_{ib}(B)$,
for any $B$ which is analytic in a disk centered at $0$ of radius $r>0$, 
e.g., a local $B$, and $b<r$. This way one can obtain, e.g.,
\be
\frac{\Vert [A, \tau_{t+ib}(B)]\Vert}{\Vert A\Vert}\leq \sum_{y\in V}
e^{2\vert t\vert \Vert \Phi\Vert_\lambda - \lambda d(x,y)}
C_{\tau_{ib}(B)}(y,0).
\ee
The remaining open question is to obtain good bounds for 
$C_{\tau_{ib}(B)}(y,0)$.

The next theorem provides an estimate on the spatial decay of correlations 
in states with a spectral gap. To state the gap condition precisely, we consider
a representation of the system on a Hilbert space $\H$. This means that there
is a representation $\pi:\A\to\B(\H)$, and a self-adjoint operator $H$ on
$\H$ such that
$$
\pi(\tau_t(A))=e^{itH}\pi(A)e^{-it H},\quad A\in \A .
$$
In the finite-system setting using a representation is, of course, merely for
convenience and not essential. We assume that $H\geq 0$ and that $\Omega\in\H$
is a vector state such that $H\Omega =0$. We say that the system has a spectral
gap in the representation if there exists $\delta >0$ such that $\spec (H) \cap
(0,\delta) =\emptyset$ and in that case the spectral gap, $\gamma$, is defined
by
$$
\gamma=\sup\{\delta > 0 \mid \spec(H) \cap (0,\delta) =\emptyset\}.
$$
Let $P_0$ denote the orthogonal projection onto $\ker H$. From now on, we will work in this
representation and simply write $A$ instead of
$\pi(A)$.

We will derive a bound for ground state correlations of the form
\be
\langle\Omega, A\tau_{ib}(B)\Omega\rangle
\ee
where $b \geq 0$ and $A$ and $B$ are local observables. 
The case $b=0$ is the standard (equal-time) correlation
function.

\begin{theorem}[Exponential Clustering]\label{thm:decay}
There exists $\mu>0$ such that for all $A\in\A_x$, $B\in\A_Y$, $x\notin Y$, 
for which
$P_0 B\Omega = P_0 B^*\Omega =0$, and $b$ sufficiently small,  
there is a constant $c(A,B)$ such that
\be
\left\vert \langle\Omega, A\tau_{ib}(B)\Omega\rangle \right\vert
\leq c(A,B)e^{-\mu d(x,Y)\left( 1 + \frac{\gamma^2 b^2}{4\mu^2d(x,Y)^2}\right)}
\, .
\label{decay}\ee
One can take
\be
c(A,B) = \Vert A\Vert \, \Vert B\Vert
\left( 1+\frac{2\vert Y\vert}{\pi}
+\frac{1}{\sqrt{\pi\mu d(x,Y)}}\right) ,
\label{cAB}\ee
and
\be
\mu = \frac{\gamma\lambda}{4 \Vert \Phi\Vert_\lambda + \gamma}\, .
\ee
The bound is valid for $0\leq \gamma b\leq 2\mu d(x,Y)$.
\end{theorem}

Usually, the distance between $x$ and $Y$, will be bounded below
by some $a>0$, so $c(A,B)$ in \eq{cAB} can be replaced by a constant
which depends only on the norms of $A$ and $B$ and the size of their
supports, but not the distance between them.
 
Note that in the case of a non-degenerate ground state, the condition on 
$B$ is equivalent to $\langle\Omega, B\Omega\rangle=0$. In that case, 
the theorem with
$b=0$ becomes
\be
\left\vert \langle\Omega, AB\Omega\rangle
-  \langle\Omega, A\Omega\rangle\, \langle\Omega, B\Omega\rangle\right\vert 
\leq c(A,B)\, e^{-\mu d(x,Y)} .
\label{zerob}\ee

For $b>0$ large, there is a trivial bound
\be
\left\vert \langle\Omega, A\tau_{ib}(B)\Omega\rangle \right\vert
\leq \Vert A\Vert \, \Vert B\Vert \, e^{-\gamma b} .
\ee
For small $b>0$, the estimate \eq{decay} can be viewed as 
a perturbation of \eq{zerob}. Often, the important observation is
that the decay estimate \eq{decay} is uniform in the imaginary time $ib$,
for $b$ in some interval whose length, however, depends on $d(x,Y)$.

\section{Proofs}\label{sec:proofs}

\subsection{Lieb-Robinson bound. Proof of Theorem \ref{thm:lr}} 

Our argument closely follows the proof of \cite[Proposition
6.2.9]{bratteli1997}, but we avoid the use of the Fourier transform in order to obtain
a generalization to arbitrary sets of vertices with a metric.

Let $A\in\A_x$ and $B\in\A$, and consider the quantities $C_{A,B}(x,t)$
and $C_B(x,t)$, defined in \eq{CAB} and \eq{CB}, respectively.

From the fundamental theorem of calculus, one has that
\be
C_{A,B}(x,t)  = C_{A,B}(x,0) \, + \, i \,
\sum_{ X \ni x} \int_0^t \, \left[\, \tau_s([\, \Phi(X),A\,]),\, B \right]\, ds\,.
\label{eq:ceqn}\ee

Several applications of the triangle inequality to (\ref{eq:ceqn})
yields
\begin{equation}
C_B(x,t) \, \leq \, C_B(x,0) \, + \, \sum_{ X \ni x} \int_0^{|t|} 
\: \sup_{A \in
  \A_x } \: \frac{ \left\| \left[\, \tau_s([\, \Phi(X),A\,]),\, B \right] \right\| }{ \| A \|} \: ds .
\end{equation}

Now, for any finite $X \subset V$ with $x \in X$, one
may write
\begin{equation} \label{eq:expphia}
\left[ \, \Phi(X), \, A \, \right] \, = \, \sum_{i=1}^{|X|} \; \:
\sum_{j_{x_i}=1,\dots,N_{x_i}^2} \: C \left( \, j_{x_1}, \ldots, j_{x_{|X|}}
  \, \right) \: \prod_{i=1}^{|X|} \, e \left( j_{x_i} \right),
\end{equation}
where each $x_i \in X$ and $e(j_{x_i})$ is a matrix unit for the algebra
$\A_{x_i}$. For each term in (\ref{eq:expphia}), expanding 
the commutator of the product yields
\begin{equation} \label{eq:basest}
 \left\| \, \left[ \,  \prod_{i=1}^{|X|} \tau_s(e(
     j_{x_i}))\, , \, B \, \right] \, \right\| \, \leq \, \sum_{y
 \in X } \, C_B \left( \, y, \, s \, \right).
\end{equation}
Combining the linearity of $\tau_s$, the basic estimate (\ref{eq:basest}), and
the fact that the coefficients in the above expansion (\ref{eq:expphia}) satisfy
\begin{equation}
\left| \, C \left( \, j_{x_1}, \ldots, j_{x_{|X|}} \, \right) \,
\right| \, \leq \, 2 \, \| \Phi(X) \| \, \| A \|,
\end{equation}
we arrive at the inequality
\begin{equation} \label{eq:cineq}
C_B(x,t) \, \leq \, C_B(x,0) \, + \, 2  \, \sum_{ X \ni x}  \, \| \Phi(X) \| \,
N^{2 \, |X|} \, \int_0^{|t|} \sum_{y \in X } C_B(y,s)
\, ds.
\end{equation}
Motivated by the expression above, we define the
quantity
\begin{equation} \label{eq:defeps}
\epsilon(x,y) \, := \, \sum_{X \ni x,y} \| \Phi(X) \| \: N^{2|X|},
\end{equation}
and rewrite (\ref{eq:cineq}) as 
\begin{equation} \label{eq:cineq2}
C_B(x,t) \, \leq \, C_B(x,0) \, + \, 2  \, \int_0^{|t|} \sum_{ y \in
 V}  \, \epsilon(x,y) \, C_B(y,s) \, ds.
\end{equation}

Iteration of (\ref{eq:cineq2}) yields 
\begin{eqnarray}
\lefteqn{C_B(x,t) \,  \leq  \, C_B(x,0) \, + \, 2 |t| \sum_{ y \in
  V}  \, \epsilon(x,y) \, C_B(y,0)}  \nonumber \\
 &&  +  \, \frac{(2|t|)^2}{2} \sum_{ y \in
  V}  \, \epsilon(x,y) \, \sum_{ y' \in
  V}  \, \epsilon(y,y') \, C_B(y',0)  \, + \nonumber \\
&&  +  \, \frac{(2|t|)^3}{3!} \sum_{ y \in
  V}  \, \epsilon(x,y)  \sum_{ y' \in
  V}  \, \epsilon(y,y') \sum_{ y'' \in
  V}  \, \epsilon(y',y'') \, C_B(y'',0) \, +
\cdots. \nonumber
\end{eqnarray}

Recall that we assumed that the interaction satisfies a bound of the form
\begin{equation}
\| \, \Phi \, \|_{\lambda} \, = \, \sup_{x \in V} \, \sum_{X \ni x}
\, |X|  \, \| \, \Phi(X) \, \| \, N^{2|X|} \, e^{ \lambda D(X)} \:
< \: \infty,
\end{equation}
for some $\lambda>0$.
We wish to prove an exponential bound on the
quantity $C_B(x,t)$. To make this bound explicit, we set 
$\epsilon_{\lambda}(x,y) := e^{ \lambda \:
  d(x,y)} \, \epsilon(x,y)$, and observe that
\begin{eqnarray} \label{eq:sumep}
\sum_{y \in V} \, \epsilon_{\lambda}(x,y) & = & \sum_{y \in V} \, \sum_{X \ni x,y} \, \| \Phi(X) \| \,
N^{2|X|} \, e^{ \lambda \, d(x,y)} \\
 & \leq  & \sum_{X \ni x} \, \sum_{y \in X}  \, \| \Phi(X) \| \,
N^{2|X|} \, e^{ \lambda \, D(X)} \: = \: \| \, \Phi \,
\|_{\lambda}. \nonumber
\end{eqnarray}

Now, returning to the iterated version of (\ref{eq:cineq2}) above, we may 
resum the upper bound, which is allowable as all the terms are
non-negative. Using the triangle inequality often, one derives 
\begin{equation}
C_B(x,t) \, \leq \, \sum_{y \in L} \,e^{- \lambda \: d(x,y)} \, f(x,y)
\, C_B(y,0) \, ,
\end{equation}
where
\begin{eqnarray}
\lefteqn{f(x,y) \, = \, \delta_{x,y} \, + \, 2 |t| \,
    \epsilon_{\lambda} (x,y) \, + \,  \frac{(2|t|)^2}{2} \, \sum_{y' \in V} \, \epsilon_{\lambda}(x, y')
  \epsilon_{\lambda} (y',y)} \nonumber \\ &&  + \, \frac{(2|t|)^3}{3!} \,  \sum_{y'' \in V}
  \sum_{y' \in V} \, \epsilon_{\lambda}(x, y'') \, \epsilon_{\lambda}(y'',y') \,
  \epsilon_{\lambda}(y',y) \, + \, \cdots \nonumber 
\end{eqnarray}

Moreover, summing over all but the first term appearing in $f$ and 
using (\ref{eq:sumep}), we see that
\begin{equation}
f(x,y) \: \leq \:  \delta_{x,y} \, + \, \left( e^{2|t| \, \|  \, \Phi
    \, \|_{\lambda}} - 1 \right).
\end{equation}
Therefore, we have proven the estimate
\begin{equation}
C_B(x,t) \, \leq \, e^{2\, |t| \, \| \Phi \|_{\lambda}} \, C_B(x,0) \,
+ \, \sum_{ \stackrel{y \in V:}{y \neq x}} \,e^{- \, \lambda \,
  d(x,y)} \, \left( e^{2 \, |t| \,   \| \, \Phi \, \|_{\lambda}} - 1 \right)\: C_B(y,0) \, ,
\end{equation}
as claimed.\hfill \qed

\subsection{Decay of correlations. Proof of Theorem \ref{thm:decay}}

It is well-known, see e.g. \cite{bratteli1997,simon1993}, 
that for interactions $\Phi \in\mathcal{B}_{\lambda}$ and $B$
a local observable the function $t \mapsto \tau_t(B)$ 
is analytic in a disk centered at the origin.

Part of our argument requires analyticity for arbitrarily large
$|t|$. Therefore, we consider entire analytic elements $B(a)$, $a>0$,
(see \cite[Proposition 2.5.22]{bratteli1987}), with the
property that $\| B(a) \| \leq \|B\|$ and $\lim_{a \to 0} \|B(a) -B \|
=0$. In this case, the function
\begin{equation} \label{eq:faz}
f_a(z) \, := \, \left\langle \Omega \, , A \, \tau_z(B(a)) \, 
\Omega \right\rangle
\end{equation}
is entire. Observe that the quantity of interest, i.e. that which appears on the
left hand side of (\ref{decay}), corresponds to $f_0( i b)$ where 
$f_0$ is as in (\ref{eq:faz}) above with $B(a) = B$ and $b>0$. The
case $b=0$ follows by a limiting argument.

Take $T > 2  b$ and denote by $\Gamma_T$ the contour in the complex plane 
which joins $-T$ to $T$ along the real axis and $T$ to $-T$ via the
semi-circle $Te^{i \theta}$ for $\theta \in [0,
\pi]$. By analyticity, we have that
\begin{equation} \label{eq:cint}
f_a( ib ) \, = \, \frac{1}{2 \pi i} \, 
\int_{ \Gamma_T} \frac{f_a(z)}{z-ib} \, dz \, . 
\end{equation}
Using the existence of a gap $\gamma >0$ and 
the assumption $P_0 B \Omega = 0$, one easily
finds that, for $\theta\in [0,\pi]$,  
\begin{equation} \label{eq:fbd}
| \, f_a( \, Te^{i \theta} \, ) \, | \, = \, \left| \, \left\langle \Omega
  \, , A \, e^{i Te^{i \theta}H} B(a) \, \Omega \right\rangle \, \right| \,
\leq \, \| A \| \, \| B \| \, e^{- \, T \, \gamma
  \, \sin \theta}.
\end{equation}
From this, we conclude that
\begin{equation} \label{eq:1bd}
|f_0( ib ) | \, = \, \lim_{a \to 0} | f_a( ib)| \, \leq \,  \, \left| \frac{1}{2 \pi i} 
\int_{-T}^T \frac{f_0(t)}{t-ib} dt \right| \, + \, \frac{ \| A \| \, \| B \|}{ \pi}
\,  \int_0^{ \pi} e^{ - T \gamma \sin( \theta)} d \theta, 
\end{equation}
where we have used that $ \| B(a) - B \| \to 0$ as $a \to 0$. 
Since the second term vanishes in the limit $T\to\infty$,
we have shown that
\begin{equation} \label{eq:2bd}
\left| \langle \Omega \, , \, A \, \tau_{ib}(B)  \, \Omega \rangle \right| \, 
\leq \, \limsup_{T \to \infty} \, 
\left| \frac{1}{2 \pi i} \int_{-T}^T \frac{f_0(t)}{t-
  ib} dt \right|,
\end{equation}
and thus, the proof has been reduced to estimating this integral 
over the real axis. 

Our estimate follows by splitting the integrand into three terms. 
Let $\alpha >0$ and write
\begin{equation} \label{eq:feqn}
f_0(t) e^{\alpha b^2}= f_0(t) e^{ - \alpha t^2} + f_0(t) 
\left( e^{\alpha b^2} - e^{ - \alpha t^2} \right).
\end{equation}
The first term above may be written so that the commuator explicitly
appears, i.e.,
\begin{equation} \label{eq:fwcom}
f_0(t) e^{- \alpha t^2} = \langle \Omega, \tau_t(B) A
\Omega \rangle \, e^{ - \alpha t^2} \, +  \, \langle \Omega, [A, \tau_t(B) ] 
\Omega \rangle \, e^{ - \alpha t^2}.
\end{equation}
Using (\ref{eq:feqn}) and (\ref{eq:fwcom}), we arrive at a 
bound of the form: 
\begin{equation} \label{eq:3bd}
 \left| \frac{1}{2\pi i} \int_{-T}^T \frac{f_0(t)}{t-
  ib} dt \right| \, \leq \, I \, + \, II \, + \, III\, , 
\end{equation}
where $I$ corresponds to the intergral containing the first term
of \eq{eq:fwcom}, $II$ contains the second term of \eq{eq:fwcom},
and $III$ the second term of \eq{eq:feqn}.
Each of these terms will be bounded separately in the limit $T \to
\infty$. 

\vspace{.5cm}

\noindent {\em The first term}: The first term appearing in the bound
on the right hand side of (\ref{eq:3bd}) is
\begin{equation} \label{eq:I}
 I \, := \, \left| \frac{1}{2\pi i} 
 \int_{-T}^T \frac{\langle \Omega, \tau_t(B) A
\Omega \rangle \, e^{ - \alpha t^2} }{t-
  ib} dt \right|.
\end{equation}
Using the spectral theorem, we may write
\begin{equation} \label{eq:st}
\langle \Omega, \tau_t(B) A \Omega \rangle \, = \, \langle e^{i t H}
B^* \Omega, A \Omega \rangle \, = \, \int_{\gamma}^{\infty} e^{-itE} d
\langle P_E B^* \Omega, A \Omega \rangle, 
\end{equation}
where $P_E$ is the projection valued spectral measure
corresponding to $H$, and we have used that $P_0 B^* \Omega =
0$. One may now write
\begin{equation} \label{eq:intst}
\int_{-T}^T \frac{\langle \Omega, \tau_t(B) A
\Omega \rangle \, e^{ - \alpha t^2} }{t-
  ib} dt \, = \, \int_{\gamma}^{\infty} \int_{-T}^T
\frac{e^{-itE} e^{- \alpha t^2}}{t-ib} \, dt \, d \langle
P_E B^* \Omega, A \Omega \rangle.
\end{equation}
The inner intergral can be estimated using Lemma \ref{lem:limit},
proven below, from which we have
\begin{equation}
\limsup_{T\to \infty} \, I \, 
\leq \, \frac{\Vert A\Vert \, \Vert B\Vert}{2}  \, 
e^{ - \frac{ \gamma^2}{ 4 \alpha}},
\end{equation}
independent of $b$.
\vspace{.5cm}

\noindent {\em The second term}: The second term is the one for which we
shall apply the Lieb-Robinson estimate in the form of \eq{beast}:
\begin{equation} \label{eq:II}
 II \, := \, \left| \frac{1}{2\pi i}  
 \int_{-T}^T \frac{\langle \Omega, [A, \tau_t(B)]
\Omega \rangle \, e^{ - \alpha t^2} }{t-
  i b} dt \right| \, \leq \, 
  \frac{1}{2\pi} \int_{- \infty}^{ \infty}
\frac{ \| \, [A, \tau_t(B)] \, \|  }{ |t|} \, e^{ - \alpha t^2} \, dt.
\end{equation}
Here it is crucial that the ``time'', $ib$, is purely imaginary.
We will now break the integral up into two regions: $ \{ t \in \mathbb{R} : |t|
\leq s \}$ where the Lieb-Robinson bound is useful and 
$ \{ t \in \mathbb{R} : |t| \geq s \}$ in which we will
use a norm estimate:
\begin{eqnarray} 
&&\frac{1}{2\pi}\int_{- \infty}^{ \infty}
\frac{ \| \, [A, \tau_t(B)] \, \|  }{ |t|} \, e^{ - \alpha t^2} \, dt
\, \nonumber\\
&&\quad \quad
\leq \,
\frac{2 \, \vert Y\vert \, \Vert A\Vert\, \Vert B\Vert}{\pi} e^{ 2 s \| \Phi \|_{\lambda} - \lambda d(x, Y)} \, + \,
\frac{\Vert A\Vert\, \Vert B\Vert}{s \, \sqrt{\pi \, \alpha}} \, e^{ - \alpha s^2}. 
\label{eq:lrbd}
\end{eqnarray}

\vspace{.5cm}

\noindent  {\em The third term}: The final term is
\begin{equation}
III \, := \,  \left| \frac{1}{2\pi i} \int_{-T}^T 
\frac{\langle \Omega, A \tau_t(B) 
\Omega \rangle \, (e^{\alpha b^2} - e^{ - \alpha t^2}) }{t-
  i b} dt \right|,
\end{equation}
which represents the penalty we incur for introducing the gaussian 
factor $e^{- \alpha t^2}$ and the scaling $e^{\alpha b^2}$. 
This term can be made small by a judicious
choice of parameters. Using the spectral theorem again, we have
\begin{equation} 
III \, = \, \left\vert\int_{\gamma}^{\infty} 
\frac{1}{2\pi i} \int_{-T}^T \frac{e^{itE} \left(e^{\alpha b^2} - e^{ - \alpha
t^2}\right)}{t-ib}
dt\, d\langle A^* \Omega, P_E B\Omega \rangle\right\vert, 
\end{equation}
By adding and substracting the appropriate limiting quantitites,
which are shown to exist in Lemma \ref{lem:limit}, the integrand
above can be rewritten as
\begin{equation}
\left(e^{\alpha b^2} e^{-Eb}- \frac{1}{2\sqrt{\pi\alpha}}
\int_0^\infty\, e^{-b \, w} \, e^{-\frac{(w-E)^2}{4\alpha}}dw\right)
+ e^{\alpha b^2} \left(F_{ib,T}(E)-e^{-Eb} \right)+ R_1 + R_2 .
\label{eq:integrand}\end{equation}
The quantities $F_{ib,T}, R_1$, and $R_2$, are defined 
in \eq{eq:FeT}, \eq{eq:R1}, and \eq{eq:R2} respectively. 
Using (\ref{eq:ftcrap}) for complex $t$, one may estimate the first term 
in \eq{eq:integrand} as 
\be
\frac{1}{2\sqrt{\pi\alpha}}
\int_{-\infty}^0\, e^{-b \, w} \, e^{-\frac{(w-E)^2}{4\alpha}}dw
\, \leq \, \frac{1}{2} e^{-\frac{\gamma^2}{4\alpha}}\, ,
\ee
for $2 \alpha b \leq \gamma$ and $E \geq \gamma$.
Using dominated convergence, one sees that 
the other terms vanish in the limit $T\to\infty$.
This implies the following bound:
\be
\limsup_{T\to\infty} III \,\leq\, \frac{1}{2}\Vert A\Vert\, \Vert B\Vert
 e^{-\frac{\gamma^2}{4\alpha}} .
\ee

The proof is completed by choosing $\alpha = \frac{ \gamma}{2s}$ and 
$s$ so that 
\begin{equation}
s \left( 2 \| \Phi \|_{\lambda} + \frac{ \gamma}{2} \right) \, 
= \, \lambda \, d(x, Y).
\end{equation}
We have that
\begin{equation}
\left| \langle \Omega, A \, \tau_{ib}(B) \Omega \rangle
  \right| \, \leq \, 
\Vert A \Vert \, \Vert B \Vert
\left(1+\frac{2 \, \vert Y\vert}{\pi}
+\frac{ 1}{ \sqrt{\pi \, \mu \, d(x,Y)}} \right)
 e^{ - \mu d( x, Y)\left(1+\frac{b^2}{s^2}\right)}\, ,
\end{equation}
where
\begin{equation}
\mu = \frac{ \lambda\gamma}{ 4 \| \Phi \|_{\lambda} + \gamma}.
\end{equation}
This proves Theorem \ref{thm:decay}. \hfill \qed

In the proof above we  used the following lemma, which is a variation
of Lemma 3.1 in \cite{hastings2005}.

\begin{lemma} \label{lem:limit} 
Let $E \in \mathbb{R}$, $\alpha >0$, and $z \in \mathbb{C}^+ := \{ z
\in \mathbb{C}: \, {\rm Im}[z] >0 \}$. One has that
\begin{equation} \label{eq:bigbad}
\lim_{T \to \infty} \, \frac{1}{2 \pi i}
\, \int_{-T}^T \, \frac{e^{iEt} \, e^{- \alpha t^2}}{t - z} \, dt \, = \, \frac{ 1}{2 \sqrt{ \pi \alpha}} \,
\int_0^{ \infty} \, e^{iwz} \, e^{ - \frac{(w-E)^2}{4 \alpha}} \, dw.
\end{equation}
Moreover, the convergence is uniform in $z\in \mathbb{C}^+$.
\end{lemma}

\noindent
{\em Proof.\/} One may prove this lemma by observing some simple
estimates on the following function:
\begin{equation}
F_{z, T}( w) \, := \, \frac{1}{2 \pi i} \, \int_{-T}^T \,
\frac{e^{i w t}}{t - z} \, dt.
\label{eq:FeT}
\end{equation}
For $w >0$, integration over the rectangular contour $\Gamma_T$ joining $-T \to T
\to T+iT \to -T +iT \to -T$ yields
\begin{equation}
\frac{1}{2 \pi i} \int_{ \Gamma_T} \, \frac{e^{i w z'}}{ z' - z} \,
dz' \, = \, e^{ iwz}
\end{equation}
for any $0<2|z|<T$. From this one may conclude that
\begin{equation} \label{eq:Fbd+}
 \left| F_{z, T}(w) - e^{iwz} \right| \, \leq \, \frac{2}{\pi} \, 
\left[ \frac{1}{ w T} \, \left( 1 - e^{- w T} \right) \, + \, e^{-
   w T} \, \right].  
\end{equation}
If $w<0$, then closing a similar rectangular contour in the lower half
plane yields
\begin{equation} \label{eq:Fbd-}
 \left| F_{ z, T}(w) \right| \, \leq \, \frac{2}{\pi} \, 
\left[ \frac{1}{ |w| T} \, \left( 1 - e^{- |w| T} \right) \, + \, e^{-
   |w| T} \, \right].    
\end{equation}

Now, as the Fourier Transform of a gaussian is another gaussian, one
easily sees that for any $E \in \mathbb{R}$ and $\alpha >0$,
\begin{equation} \label{eq:ftcrap} e^{iEt} \, e^{ - \alpha t^2} \, = \, \frac{1}{2 \sqrt{ \pi \alpha}} \,
\int_{- \infty}^{ \infty} \, e^{i w t} \, e^{- \frac{(w-E)^2}{4
    \alpha}} \, dw.
\end{equation}
We may now write the prelimit quantity on the left hand side of
(\ref{eq:bigbad}) as the desired limit and two remainder terms, i.e.,
\begin{eqnarray}
\frac{1}{2 \pi i} \, \int_{-T}^T \, \frac{e^{iEt} \, e^{- \alpha t^2}}{t - z}
 \, dt & = & \frac{1}{2 \sqrt{ \pi \alpha}} \,
\int_{- \infty}^{ \infty} \, e^{ - \frac{(w-E)^2}{4 \alpha}} \, F_{
 z, T}(w) \, dw \nonumber \\
& = & \frac{1}{2 \sqrt{ \pi \alpha}} \,
\int_0^{ \infty} \, e^{i w z} \, e^{ - \frac{(w-E)^2}{4 \alpha}} \, dw \, + \, R_1
\, + \, R_2.
\end{eqnarray}
Using the estimates (\ref{eq:Fbd+}) and (\ref{eq:Fbd-}), one may apply
dominated convergence to conclude that both the remainders
\begin{equation} \label{eq:R1}
R_1 \, := \, \frac{1}{2 \sqrt{ \pi \alpha}} \,
\int_0^{ \infty} \, e^{ - \frac{(w-E)^2}{4 \alpha}} \, \left(F_{
    z, T}(w) - e^{iwz} \right) \, dw
\end{equation}
and
\begin{equation} \label{eq:R2}
R_2 \, := \, \frac{1}{2 \sqrt{ \pi \alpha}} \,
\int_{- \infty}^0 \, e^{ - \frac{(w-E)^2}{4 \alpha}} \, F_{
  z, T}(w) \, dw
\end{equation}
vanish in the limit $T \to \infty$.
\hfill \qed

\section{Applications and Generalizations}\label{sec:appgen}

For many applications it is of interest to consider the thermodynamic
limit. Our clustering theorem allows for two approaches to the thermodynamic
limit, each with its own merits. In the first approach, one applies the theorem
to finite systems and obtains estimates that are uniform in the size
of the systems.
This may require a careful choice of boundary conditions. The bounds then
carry over automatically to any state which is a thermodynamic limit of 
finite volume states for which the estimates are obtained. This is the most
straightforward way to proceed in cases where the finite volume ground states are 
unique, weak$^*$ convergent and with a uniform lower bound on the spectral gap. 

In the second approach one focuses directly on the infinite system. This requires a
proof of a spectral gap for the infinite system, but may avoid difficulties associated
with boundary states which may obscure the spectral gap for finite volumes.

Another method for dealing with boundary states is to generalize Theorem
\ref{thm:decay} to systems with quasi-degenerate ground states. As long as the
quasi-degenerate states remain separated from the excitation spectrum by a 
non-vanishing gap, this is a straightforward generalization, but the condition 
$P_0 B \Omega=0$, where now $P_0$ is the spectral projection on all states
below the gap, may be a bit tricky to verify.

Another rather obvious generalization of our results is to systems of fermions
on $(V,d)$. For even interactions (only products of an even number of creation
and annihilation operators occur), and even observables, all results carry over without
change. If $A$ and $B$ are both odd observables, one gets a Lieb-Robinson bound for
the anticommutator instead of the commutator.

\vspace{.5cm}

\noindent
{\it Acknowledgements:}
Based on work supported by the U.S. National Science
Foundation under Grant \# DMS-0303316. We thank M. B. Hastings and T. Koma for
useful comments on earlier versions of the manuscript. We thank the referee
for pointing out an error in a previous draft.

\end{document}